\documentclass[12pt]{iopart}
\usepackage{iopams} 
\usepackage{graphics}
\usepackage{graphicx}
\usepackage{pstricks}
\usepackage{pst-node}
\usepackage{multido}
\usepackage{pst-plot}
\usepackage{pst-coil}
\usepackage{psfrag}
\usepackage{verbatim}

\newcommand{\bra}[1]{\ensuremath{\langle#1|}}
\newcommand{\ket}[1]{\ensuremath{|#1\rangle}}

\newcommand{\openone}{\mathbb I}
\def\sgen{stochastic finite-state generator}
\def\vNgen{von Neumann finite-state generator}
\def\HQM{HQMM}

\def\Abet {\mathcal{A}}
\newcommand{\svector}[1]{\ket{ {#1} }}
\newcommand{\Prob}      {\mathrm{Pr}}
\newcommand{\pangle}      {\phi}
\newcommand{\basis}      {\epsilon}
\newcommand{\symb} {s}

\newcommand{\stat}{\star}

\newcommand{\rank}{\textrm{rank}}

\usepackage[latin1]{inputenc}
\usepackage{amsfonts}
\usepackage{amssymb}

\begin{document}

\title{Hidden Quantum Markov Models and non-adaptive read-out of many-body states}
\author{Alex Monras}
\address{Dipartemento di Matematica e Informatica, Universit\`a degli Studi di Salerno, Fisciano (SA), Italy\\
The School of Physics and Astronomy, University of Leeds, Leeds LS2 9JT, United Kingdom 
}
\author{Almut Beige}
\address{The School of Physics and Astronomy, University of Leeds, Leeds LS2 9JT, United Kingdom}
\author{Karoline Wiesner}
\address{School of Mathematics, Centre for Complexity Sciences, University of Bristol, Bristol BS8 1TW, United Kingdom}
%\ead{k.wiesner@bristol.ac.uk}
\date{\today}

\begin{abstract}
Stochastic finite-state generators are compressed descriptions of infinite time series. Alternatively, compressed descriptions are given by quantum finite-state generators [K. Wiesner and J. P. Crutchfield, Physica D {\bf
237}, 1173 (2008)]. These are based on repeated von Neumann measurements on a quantum dynamical system.
Here we generalise the quantum finite-state generators by replacing the von Neumann projections by stochastic quantum operations. In this way we assure that any time series with a stochastic compressed description has a compressed quantum description. Moreover, we establish a link between our stochastic generators and the sequential readout of many-body states with translationally-invariant matrix product state representations. As an example, we consider the non-adaptive read-out of 1D cluster states. This is shown to be equivalent to a Hidden Quantum Model with two internal states, providing insight on the inherent complexity of the process. Finally, it is proven by example that the quantum description can have a higher degree of compression than the classical stochastic one. 
\end{abstract}

\pacs{05.40.-a, 03.67.Ac, 89.70.Eg, 89.70.+c}
%\submitto{\NJP} 
\bibliographystyle{unsrt}

\maketitle

\section{Introduction} \label{sec.intro}

The physical laws underlying quantum computation are a mixed blessing. On the one hand, a growing body of theoretical results suggests that a computational device whose components are directly governed by quantum physics may be considerably more powerful than its classical counterpart. As proof, several quantum algorithms have been constructed. The most celebrated ones are the factoring algorithm by Shor \cite{shor:94}, suggesting an exponential speed-up over classical algorithms, and the data-base search algorithm by Grover \cite{grover:96}, proving a linear speed-up. They are quantum versions of algorithms that improve the efficiency of algorithms usually implemented on classical Turing machines. 

Contrary to this progress on the theoretical side, building a quantum computer in practice faces many hurdles. Implementations of quantum Turing machines must maintain high degrees of internal coherence and insulation to outer disturbances during operation. Thus, the current implementations remain with highly finite systems, much more of the character of finite-state machines than Turing machines \cite{knill:00}. In the yet to be completed hierarchy of quantum computational architectures quantum Turing machines are decidedly more powerful than finite-state machines. However, because of the ``finiteness'' of current implementations of quantum algorithms the lower part of this hierarchy and its potential computational power are of great interest. 

At this point, a look at information creation in dynamical systems turns out to be useful. Recently, one of the authors introduced a synthesis of quantum dynamics, information creation, and computing \cite{wiesner_computation_2008, crutchfield_intrinsic_2008}. This lead to methods useful for analysing how quantum processes store and manipulate information. A {\em quantum process} was defined as a probability distribution over measurement outcomes of a repeatedly measured quantum system, formally represented as a process language. A measurement-based quantum-computation hierarchy was developed, providing insights into the information processing in quantum dynamical systems. This computational hierarchy allows for a classification of quantum dynamical systems in terms of the process languages they generate. Following this ansatz, we concentrate here on a characterisation of generators of process languages. The characterisation of recognisers is analogous (up to some exceptions \cite{wiesner_computation_2008}) but will not be addressed here.

The quantum finite-state machines introduced by Wiesner and Crutchfield in
2008 \cite{wiesner_computation_2008} generate bit sequences by applying a
projective (von Neumann) measurement repeatedly to an otherwise dynamically
evolving quantum system. In the following, we therefore refer to them as {\em
quantum von Neumann finite-state generators}. However, there are regular
languages which cannot be generated by a von Neumann finite-state machine
\cite{wiesner_computation_2008}. An example is an alternating binary sequence
with different probabilities for the two binary symbols
\cite{wiesner_computation_2008}. Here we point out that it is possible to
overcome this problem by replacing the projective measurements of von Neumann
finite-state machines by stochastic quantum operations, \emph{i.e.},
generalized measurements described in the quantum operation formalism. We call
the resulting stochastic processes \emph{Hidden Quantum Markov Models} (\HQM).

Quantum operations are the most general transformations that a quantum state can undergo. This includes the possibility of the state being (partially) measured and the  possibility of resetting or transforming the system conditional on the measurement outcome \cite{kraus_operations_1974}. Contrary to projective measurements, the state of the system after a measurement may depend not only on the respective measurement outcome. In general, it may retain information about the state of the system prior to the operation and hence also information about previous measurement outcomes. Thus, the system acquires a memory which may be anything from classical to quantum. As a consequence, any process language which can be generated by a classical stochastic finite-state machine can now be generated by a quantum finite-state machine. Secondly, any iterative quantum process can now be represented as a quantum finite-state machine. Finally, we show by example that there are process languages which are generated with fewer resources using quantum finite-state machines compared to classical ones.

There are six sections in this paper. Section~\ref{sec.stocproc} reviews stochastic processes, their formal representation as process languages, and stochastic finite-state machines as generators of such process languages. This is followed by a review of quantum processes and {\vNgen}s in Section~\ref{sec.qproc}. Section~\ref{sec.Kgen} introduces {\HQM}s as computation-theoretic representations of general finite-dimensional quantum operations and presents an information-theoretic analysis of quantum systems under generalised measurements. Section~\ref{sec:Cluster-state} shows that the non-adaptive read-out of a 1D cluster state generates the same process languages as a repeatedly measured dynamical open quantum two-level system. It therefore constitutes an example of a \HQM. Finally, we summarize our results and draw attention to some open questions in Section \ref{sec:conclusions}.

\section{Stochastic Processes} \label{sec.stocproc}
Consider the temporal evolution of some natural system. The evolution is monitored by a series of measurements---numbers sequentially registered. Considering the numbers to be discrete, each such measurement can be taken as a discrete random variable $X_i$. The probability distribution $\Prob(X_{-\infty}^{+\infty})$ over bi-infinite sequences $X_{-\infty}^{+\infty}$ of these random variables is what we refer to as a \emph{stochastic process}. Such a process is a complete description of a system's behaviour. An important question for understanding the structure of natural systems therefore is what kinds of stochastic processes there are. This question can be studied using stochastic finite-state generators which are a classification tool for stochastic processes.

The distribution over sequences of random variables as a representation of a
stochastic process is merely an enumeration of probabilities. It is not a very
practical representation, let alone does it allow for any insights. Thus, it
is desirable to find a more compact representation of a process than just the
probability distribution over sequences. 

Among all possible stochastic processes, several models have proven successful in order to model a wide range of practical situations. One such class of processes are the so-called Hidden Markov Models (HMM from now on). There are several possible ways to define a HMM. The most common among mathematicians is as a stochastic process $\textbf{S}$ for which every symbol $s_n$ is generated conditionally from a Markovian process $\textbf{X}$, with probability $\Prob(s_n|x_n)$. The Markov process $\textbf{X}$ goes unobserved. This definition is called the Moore HMM. An equivalent definition, more common in the computer science literature is that of a Mealy HMM, which can be formulated as the process generated by a stochastic finite-state generator, and will be the one considered in the present work.
Stochastic finite-state generators differ from finite-state machines used for
recognition of regular languages in two ways. They have probabilistic
transitions, similar to Paz's probabilistic automata
\cite{paz_introduction_1971}. And they generate sequences instead of
recognising them. In the following, we use the minimal number of internal
states that a device needs in order to generate a stochastic language as a
measure of complexity \cite{grassberger_towardquantitative_1986, crutchfield_inferring_1989}. This quantification of complexity is complimentary to the notion of algorithmic complexity~\cite{cover_elements_2006,li_introduction_1997}. 

\subsection{Stochastic finite-state generators and hidden Markov models}

Let us recall the definition of a \sgen~from Ref.~\cite{wiesner_computation_2008}. A \emph{\sgen} is a tuple $(S,\Abet, {\bf T})$ where $S$ is a finite set of states, 
$\Abet=\{s\}$ is a countable set of symbols, output alphabet, and ${\bf T} = \{T_s:s\in\Abet\}$ are square substochastic matrices of dimension $|S|$, such that $\sum_s T_s$ is stochastic. The system undergoes transitions between internal states and every transition has an associated output symbol. The generator is fully characterized when all conditional probabilities $\Prob(s;j|j_0)$ are given, that is, the probability of yielding symbol $s\in\Abet$ and going to state $j$ conditional on the machine being in state $j_0$. These probabilities are summarized by the set of substochastic matrices $T_s$, where $[T_s]_{j\,j_0}=\Prob(s;j|j_0)$~\footnote{In \cite{wiesner_computation_2008} the authors used row vectors and right multiplication. Here we use column vectors and left multiplication in accordance with standard notation in quantum information.}. The sum $\sum_{s
\in\Abet}T_s$ is a stochastic matrix which describes the Markovian evolution of the state space when the output symbols are disregarded. These automata can be represented by a directed graph where nodes correspond to states and directed edges correspond to transitions with non-zero probability (see Ref.~\cite{wiesner_computation_2008} for details and Fig.~\ref{fig.even} for an example of such a representation). 

The computation of word probabilities is easily done in terms of the matrices $T_s$. The conditional probability of word $\textbf{s}=(s_1,\ldots,s_n)$ given an initial state $j_0$ is
\begin{equation}
	\Prob(\textbf{s}|j_0)=\sum_{j_1,j_2,\ldots j_{n}}\Prob(s_n;j_n|j_{n-1})\cdots\Prob(s_2;j_2|j_1)\Prob(s_1;j_1|j_0)
\end{equation}
which can be written as
\begin{equation}
	\Prob(\textbf{s}|j_0)=\sum_{j_n}\left[T_{\textbf{s}}\right]_{j_n\,j_0}~,
\end{equation}
where we have defined $T_{\textbf{s}}=T_{s_n}\cdots T_{s_2}T_{s_1}$. Finally, the state of the machine can be a statistical mixture of all possible internal states, which is conveniently described by a vector $\ket \pi$. This can be used as a prior distribution when the initial state is uncertain. The stable stochastic process is obtained from the steady state of the machine, defined as the eigenvector $\ket{\pi^\stat}$ with eigenvalue 1 of $T=\sum_{s}T_s$, namely
\begin{equation}\label{eq:ssSG}
	\ket{\pi^\stat}=\sum_{s\in\Abet} T_s \ket{\pi^\stat}~.
\end{equation}
Finally, word probabilities can be computed by defining the vector $\ket{1}= (1,1,\ldots,1)^T$,
\begin{equation}\label{eq:wpSG}
	\Prob(\textbf s)=\bra{1} T_{s_n}\cdots T_{s_2}T_{s_1} \ket{\pi^\stat}~.
\end{equation}

Given a stochastic finite-state generator and a prior probability distribution $\ket{\pi_0}$ of the machine state, all word probabilities are established and a HMM is uniquely defined.

\subsection{Example of a stochastic finite-state generator} \label{Even}

\begin{figure}
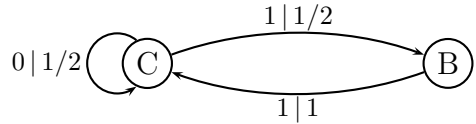
  
\begin{center}
% \resizebox{2.7in}{!}{\includegraphics{even-graph.eps}}
	\pspicture(1,2)(8,4)
	\psset{radius=.5,labelsep=0.0}
	%\psgrid
	\cnodeput(7,3){right}{B}
	\cnodeput(3,3){left}{C}

	\nccurve[angleA=200,angleB=-20]{->}{right}{left}
	\naput{\footnotesize $1\,|\,1$}
	
	\nccircle[angleA=90]{->}{left}{.4}
	%\nccircle[]{->}{left}{.4}
	\nbput{\footnotesize$0\,|\,1/2\,$}
	
	\nccurve[angleA=20,angleB=160]{->}{left}{right}
	\naput{\footnotesize $1\,|\,1/2$}
	\endpspicture
\end{center}
\caption{A deterministic generator of the Even-Process: Blocks of an even
  number of $1$s are separated by $0$s. Only the asymptotically recurrent
  states are shown.
  Edges are labeled $p \,| \, \symb$, where $\symb \in \Abet$ and $\Prob(s;i|j) = [T_s]_{ij}$. 
  }
\label{fig.even}
\end{figure}

As an example we give the \sgen~for the {\em Even-Process} whose language consists of blocks of even numbers of $1$'s bounded by $0$'s. The substochastic transition matrices are
\begin{eqnarray} \label{eq.ep}
T_0 = \left(\begin{array}{cc} \frac{1}{2} & 0 \\ 0 & 0
  \end{array}\right) ~~\mathrm{and}~~
T_1 = \left(\begin{array}{ccc} 0 & 1 \\ \frac{1}{2} & 0
  \end{array}\right) ~.
\end{eqnarray}
Again, notice that left-multiplication is used. The corresponding graph is shown in Fig.~\ref{fig.even}. 
The set of irreducible forbidden words is countably infinite \cite{weis73}:
$\mathcal{F} = \{ 01^{2l+1}0: l = 0, 1, 2, \ldots \}$. As a
consequence the words in the Even-Process have a kind of infinite
correlation: the ``evenness'' of the length of $1$-blocks is respected
over arbitrarily long words. This makes the Even-Process effectively
non-finite: As long as a sequence of $1$'s is produced, memory of the
sequence persists \cite{kitchens_symbolic_1997}.   

\section{Quantum Processes} \label{sec.qproc}

We now turn to quantum systems and the process languages they generate. As
with stochastic processes, we assume that the evolution of a quantum system is
recorded as a sequence of measurement outcomes. Just like before, the distribution over sequences of these random variables is a
\emph{process}. A computation-theoretic representation of a process generated
by a quantum systems is the quantum finite-state machine introduced by one of
the authors \cite{wiesner_computation_2008}. These machines, reviewed below,
are based on a von Neumann measurement. In order to
distinguish them from the more general Hidden Quantum Models introduced later on, we will
call them in the following {\emph \vNgen s}.

\subsection{Quantum {\vNgen}s} \label{sec.vNgen}

Like stochastic generators, a quantum {\vNgen}, as introduced in \cite{wiesner_computation_2008}, outputs a symbol $\symb \in \Abet $ and updates its internal state $\psi$. Now $\psi$ belongs to an $|S|$-dimensional Hilbert space $H$ instead of belonging to a set $S$, and is consistently denoted throughout the quantum information literature by $\ket{\psi}$. The classical states $j\in S$ correspond to an orthonormal basis in $H$, $\{\ket{\psi_j}\}$. Properly normalized linear combinations of $\ket{\psi_j}$ are also allowed as state vectors. Symbols are obtained by measuring the quantum system. For the von Neumann generators only von Neumann measurements are considered, which are described by a complete set ${\bf P}$ of projection operators, 
\begin{eqnarray}
\mathbf{P} &=& \{ P_s:  s\in\Abet~{\rm and} ~ \sum_s P_s = \openone \} ~\, . 
\end{eqnarray}
Between measurements, a unitary transformation $U$ is applied to the system. The overall effect can be described by the operators $T_s=P_s U$. Starting with the state $\svector{\psi_{t}}$ at time $t$  symbol $s$ is put out and $\ket {\psi_{t+1}(s)} = P_s U \ket{\psi_t}$ is the (unnormalised) state
 at the next time step. In general, $\ket{\psi(\textbf{s})}$ denotes the resulting state after starting with $\ket{\psi}$ and obtaining measurement outcomes $\textbf{s}$. The unnormalized state $\ket{\psi(\textbf{s})}$ can be expressed as
\begin{equation}
	\ket{\psi(\textbf{s})}=T_{\textbf{s}}\ket\psi~,
\end{equation}
where again $T_{\textbf{s}}=T_{s_n}\cdots T_{s_2}T_{s_1}$. The probability of obtaining the word $\textbf s$ from state $\ket\psi$ is given by
\begin{equation}
	\Prob(\textbf{s})=\|\ket{\psi(\textbf{s})}\|^2=\bra\psi T_{\textbf{s}}^\dagger T_{\textbf{s}}\ket\psi~.
\end{equation}

\subsection{Density matrix representation} \label{densi}

The above formulation of a \vNgen~is developed in analogy to a \sgen. A more convenient way, and at the same time more general, is a formulation in terms of density matrices~\cite{wiesner_computation_2008, nielsen_quantum_2000}. This allows to describe the most general state of the system, accounting for classical uncertainties as well as quantum indeterminacy. A state $\ket\psi$ is then described by density matrix $\rho=\ket\psi\bra\psi$. The action of $T_{\textbf{s}}$ on $\ket\psi$ is replaced by
\begin{equation}
	\mathcal T_{\textbf{s}}\rho=T_{\textbf{s}}\ket\psi\bra\psi T_{\textbf{s}}^\dagger~.
\end{equation}
Here $\mathcal T_{\textbf{s}}$ is a linear transformation on $\rho$, usually called a \emph{superoperator}. The probability of obtaining word $\textbf{s}$ then reads
\begin{equation}
	\Prob(\textbf{s})=\tr[\mathcal T_\textbf{s}\rho]~.
\end{equation}
The density operator formalism will allow for a powerful generalization of the \vNgen.

\subsection{Example of a von Neumann finite-state generator}

An example of a \vNgen~is shown in Fig.~\ref{fig.qep}. It generates the Even-Process language we are already familiar with. It consists of blocks of even numbers of $1$'s bounded by $0$'s. It's transition matrices $T_s$ can be constructed from the following projection operators $P_s$ and the unitary matrix $U$~\cite{wiesner_computation_2008},
\begin{eqnarray}
 P_0 = \left(
	\begin{array}{ccc}
		1 & 0 & 0\\
		0 & 0 & 0\\
		0 & 0 & 0
	\end{array}
	\right) ,~
P_1 = \left(
	\begin{array}{ccc}
		0 & 0 & 0\\
		0 & 1 & 0\\
		0 & 0 & 1
	\end{array}
	\right)  , ~
U = \left(
	\begin{array}{ccc}
		\frac{1}{\sqrt{2}} & 0 & -\frac{1}{\sqrt{2}}
		\\ \frac{1}{\sqrt{2}} &0 & \frac{1}{\sqrt{2}}  
		\\ 0 & -1 & 0
	\end{array}
	\right)~.
\label{eq.qep}
\end{eqnarray}
The corresponding directed graph is shown in Fig.~\ref{fig.qep}. This generator consists of 3 internal quantum states. This is the smallest dimension needed to generate the Even Process with this kind of generators.

\begin{figure}[t]
\begin{center}
	\psset{radius=.5,labelsep=0.0,unit=1.5cm}
	\pspicture(2,2)(6,6)
	%\psgrid
	\cnodeput(4,4.5){up}{A}
	\cnodeput(5,3){right}{B}
	\cnodeput(3,3){left}{C}
	
	\nccurve[angleA=-25,angleB=90]{->}{up}{right}
	\naput{\footnotesize$1\,|\,1/\sqrt{2}$}

	\nccircle[]{->}{up}{.35}
	\nbput{\footnotesize $0\,|\,1/\sqrt{2}$}

	\nccurve[angleA=90,angleB=205]{->}{left}{up}
	\naput{\footnotesize$0\,|\,-1/\sqrt{2}$}

	\nccurve[angleA=205,angleB=-25]{->}{right}{left}
	\naput{\footnotesize $1\,|\,-1$}

	\nccurve[angleA=25,angleB=155]{->}{left}{right}
	\naput{\footnotesize $1\,|\,1/\sqrt{2}$}
	\endpspicture

\caption{\vNgen ~for the Quantum Even-Process.} \label{fig.qep}
\end{center}
\end{figure}

\subsection{More general quantum processes}

Notice that the \vNgen~above has three states which is more than the stochastic generator in Section \ref{Even} required. This means, the quantum representation is less compressed than the classical one. This is rather counterintuitive. As we shall see below, the reason for this is the restriction to unitary transformations $U$ and projective measurements $P_s$. No room is left for interaction with auxiliary quantum systems, or the resetting of parts of the system. However, physical systems with such couplings -- and many quantum technology experiments relying on them -- can be relatively simple.  Hence realistic measurements are often  imperfect and noisy, where the von Neumann measurements are an idealization. We will see that extending the formalism to density operators will allow for, not only describing more realistic processes, but also encompass the whole class of classical generators.

\section{Hidden Quantum Models}
\label{sec.Kgen}
In this section we define the Hidden Quantum Markov Model (\HQM), based on the
notion of quantum operation. The spirit underlying our definition follows
exactly the same lines as the definition of a HMM. In a HMM (\HQM) there is an
underlying Markovian (Quantum) process which rules the probability of the
symbols generated. We will see how word probabilities and the steady state for
{\HQM}s are obtained in full analogy to HMMs.

\subsection{Quantum operations} \label{sec:operations} \label{Krausops}

This section is a review of the theory of the so-called quantum operations, introduced in the seminal papers~\cite{kraus_operations_1974, hellwig_pure_1969, hellwig_operations_1970} and for which several reviews are available (see, for example~\cite{nielsen_quantum_2000, bengtsson_geometry_2006}).

A \emph{quantum operation} $\mathcal K$ is a completely positive (CP) trace non-increasing linear map on the space of density operators (CP map in short). A theorem due to Stinespring~\cite{nielsen_quantum_2000} shows that every CP map admits a representation of the form
\begin{equation}
	\label{eq:operator-sum-representation}
	\mathcal K\rho=\sum_i K_i \rho K_i^\dagger~,
\end{equation}
where $K_i$ are linear operators acting on Hilbert space $H$, often called Kraus operators~\cite{bengtsson_geometry_2006}. Eq.~(ref{eq:operator-sum-representation}) is called the operator-sum representation of $\mathcal K$. Conversely, every map of the form (\ref{eq:operator-sum-representation}) is completely positive. It is worth stressing that the operator-sum representation is in general not unique, thus the physical significance of the Kraus operators is not always straightforward. The trace non-increasing condition naturally reads
\begin{equation}
	\sum_iK_i^\dagger K_i\leq \openone~.
\end{equation}

Quantum operations can be classified according to different criteria. They can be trace preserving $\sum_i K_i^\dagger K_i=\openone$ or trace-decreasing $\sum_i K_i^\dagger K_i<\openone$. In the former case, they map density operators to density operators, and are often regarded as \emph{quantum channels}, or more technically, CPTP maps. If they are trace-decreasing they usually are a piece of a larger object, \emph{i.e.}, a complete set of trace-decreasing operations $\{\mathcal K_s\}$, forming a \emph{stochastic quantum operation}. This corresponds to the most general transformation that a quantum state can undergo, which generates a \emph{symbol} or measurement outcome, in addition to the final state of the system. A 
stochastic quantum operation has to fulfill the condition that $\mathcal K=\sum_s \mathcal K_s$ is trace-preserving (a quantum channel), for the final state of the system has to be a properly normalized density operator when the generated symbol is discarded. The notion of stochastic quantum operation encompasses anyhting from unitary transformations  ($\mathcal K_s\,\cdot=U\cdot U^\dagger$) to von Neumann measurements ($\mathcal K_s\,\cdot=P_s\cdot P_s$), as well as generalized measurements such as POVMs.\footnote{A remark about POVMs is in order. A POVM is a powerful tool to compute the outcome probabilities of any possible quantum mechanical measurement, but it does not provide information about the resulting state after the measurement. Quantum operations do provide this information, while the POVM element can always be recovered from the operator-sum representation.}

Other possible classifications are, whether they are \emph{pure} operations or not. A pure operation maps pure states onto pure states, which means that there exists an operator-sum representation which consists of only one Kraus operator. The action of a pure operation can be tracked by a pure state whereas in general, superoperators can only be understood as acting on density operators. Further classification may distinguish between unitary and non-unitary operations. In particular, a unitary operation is trace-preserving and pure. Apart from the transformations enumerated so far, quantum operations can easily describe the physical process of coupling the system to an ancilla and measuring the ancilla, performing a transformation on the quantum system itself. The latter is dependent on a previous outcome and possibly some random variable obtained by completely external means. Some examples of these different operations will appear throughout the remainder of the paper.

Let us go back to (trace-decreasing) stochastic quantum operations. These correspond to operations that cannot be implemented with unit probability. They often represent the action of a quantum measurement on a system, for a specific outcome $s$. The probability of successfully implementing a given operation $\mathcal K_s$ on quantum state $\rho$ is given by 
\begin{equation}
	\Prob(s)=\tr[\mathcal K_s\rho]~.
\end{equation}
We can assign a set of quantum operations $\{\mathcal K_s\}$ to a set of outcomes $\Abet=\{s\}$, such that each outcome occurs with probability $\Prob(s)$ and the conditional state after outcome $s$ is
\begin{equation}
	\rho_\symb =\frac{\mathcal K_\symb  \rho}{\Prob(\symb)}~.
\end{equation}
Let $K_i(s)$ be the Kraus operators of an operator-sum representation of $\mathcal K_s$. Unit probability $\sum_\symb \Prob(\symb)=1$ is guaranteed by the fact that $\sum_s \mathcal K_s$ is trace-preserving, \emph{i.e.},
\begin{equation}
\label{eq.o-sum}
	\sum_{\symb,i}K_{i}^\dagger(\symb) K_{i}(\symb)=\openone~.
\end{equation}
which shows that the complete set of Kraus operators (running indices $i$ and $s$) form a quantum channel. This channel corresponds to performing the measurement (or stochastic channel $\{\mathcal K_s\}$) and forgetting the outcome, since the resulting state is
\begin{equation}
	\tilde\rho=\sum_\symb \Prob(\symb)\rho_\symb =\sum_{\symb,i}K_{i}(\symb)\rho K_{i}^\dagger(\symb)~.
\end{equation}
Finally, every trace-preserving operation has a steady state, defined as the eigenvector (understood in terms of superoperators as linear transformations and density operators as vectors) with eigenvalue 1, $\rho^\stat=\mathcal K\rho^\stat$. Analogously, the steady state of a stochastic quantum operation is $\rho^\stat=\sum_s \mathcal K_s \rho^\stat$.

\subsection{Definition of the Hidden Quantum Model}

Notice in Table~\ref{tab:analogy} the analogy between the structure of stochastic generators and the notion of stochastic quantum operation~\cite{bengtsson_geometry_2006}. Every piece in the classical formalism has a direct counterpart in the quantum formalism. It is thus natural to define a quantum stochastic generator following this analogy.\\*

\begin{table}\flushright
\begin{tabular}{| r | c | c |}
	\hline
	 & \emph{HMM} & \emph{\HQM}\\
	\hline
	State & $\ket{\pi}$ & $\rho$\\
	\hline
	Transitions & Substochastic $\{T_s\}$ & Trace-decreasing $\{\mathcal K_s\}$\\
	\hline
	Forgetful transition	& Stochastic $T=\sum_s T_s$& Trace-preserving $\mathcal K=\sum_s \mathcal K_s$\\
	\hline
	Steady state & $\ket{\pi^\stat}=T\ket{\pi^\stat}$ & $\rho^\stat=\mathcal K\rho^\stat$\\
	\hline 
	$\Prob(\textbf s)=$ & $\bra 1 T_{s_n}\cdots T_{s_2}T_{s_1} \ket{\pi^\stat}$&$(\openone| \mathcal K_{s_n}\cdots \mathcal K_{s_2}\mathcal K_{s_1} |\rho^\stat)$\\
	\hline
\end{tabular}
\caption{\label{tab:analogy} Analogy between the classical finite-state generator and the stochastic quantum operation. The mapping is straightforward by replacing probability vectors by density operators, transition matrices with CP-maps, of which stochastic matrices correspond to trace-preserving maps and substochastic corresponds to trace-decreasing. Finally, taking the inner product between the final state $T_{s_n}\cdots T_{s_2}T_{s_1}\ket{\pi^\stat}$ and $\bra 1$, respectively 
is equivalent to taking the Hilbert-Schmidt inner product between vectors $\mathcal K_{s_n}\cdots\mathcal K_{s_2}\mathcal K_{s_1}|\rho)$ and the identity $(\openone|$.}
\end{table}

\noindent\textbf{Definition:} A \emph{Hidden Quantum Markov Model} (\HQM) is a $d$-level quantum system $\rho$ together with a set of quantum operations $\mathcal K_\symb$ such that $\sum_\symb \mathcal K_\symb $ is trace-preserving. At every time step a symbol is generated with probability $\Prob(s)=\tr [\mathcal K_s\rho]$ and the state vector is updated to $\rho_s=\mathcal K_s\rho/\Prob(s)$.\\* 

The definition of quantum generators based on operations is useful in several ways,
\begin{itemize}
	\item The $\mathcal K_\symb $ quantum operations play, in quantum
mechanics, the role of the substochastic matrices $T_\symb$ for the SGs. They
provide a description of the whole process (state transitions as well as
probabilities). As we will see in Sec.~\ref{properties}, defining {\HQM}s in terms of quantum operations easily accommodates SGs in a straightforward way. This is a highly desired and natural feature in the sense that quantum devices are often expected to generalize classical devices.
	\item Orthodox projective measurements are ideal and often the hardest to realize. Instead, realistic measurements are often performed by coupling the quantum system of interest to some auxiliary system, or \emph{ancilla}, which is later measured. The whole process can then be regarded as a generalized measurement, described by a POVM. The extension of quantum generators to {\HQM}s is well-suited to accommodate models for such processes, arising naturally in the study of quantum systems, and hence providing a natural framework for modeling the generated stochastic processes.
	\item We will show an example case where a stochastic process needs 3
states to be generated classically, whereas only 2 states suffice when it is
generated quantum-mechanically. This improvement suggests that an increase in
efficiency may be possible, by harvesting the power of quantum mechanics in
the task of process generators. Also, {\HQM}s may provide a more efficient means to model complex stochastic processes. Whether this improvement is significant or not for highly complex processes is left as an open question.
\end{itemize}

\subsection{Properties} \label{properties}

We now derive some properties of the \HQM ~and provide prescriptions for generating them from classical processes.\\*

\noindent \textbf{Theorem:} For every SG with $d$ internal states there exists a \HQM ~with the same number of internal states which generates the same stochastic process.\\
\noindent \emph{Proof:} We prove the theorem by construction. We first provide a prescription for constructing the \HQM ~and then show that the word probabilities are consistent with the $T$ matrices.

Let the internal states of the system be represented by $\{\rho_i=\ket{i}\bra{i}\}$. The transition from state $\rho_j$ to state $\rho_i$ together with outcome $\symb$ has to occur with probability $\Prob(s;i|j)=[T_\symb]_{ij}$. For every symbol $s$ we define a quantum operation $\mathcal K_s$ with operator sum representation
\begin{equation}
	\mathcal K_s\rho=\sum_{ij} K_\symb ^{i,j}\rho K_\symb ^{\dagger\,i,j},\qquad K_\symb ^{i,j}=\sqrt{[T_\symb]_{ij}}\ket{i}\bra{j}~.
\end{equation}
Thus we achieve
\begin{equation}
	\mathcal K_\symb \rho_j=\sum_i [T_s]_{ij}\rho_i~,
\end{equation}
and the probabilities are,
\begin{equation}
	\Prob(s|j)=\tr [\mathcal K_s\rho_j]=\sum_i [T_s]_{ij}=\sum_i \Prob(s;i|j)~.
\end{equation}
Namely, the probability of outcome $s$ given initial state $j$, $\sum_i[T_s]_{ij}$ is consistent with our formulation. In addition, the final state undergoes the standard stochastic evolution according to the transition matrix $[T_\symb]_{ij}$~$\blacksquare$\\*

Notice that this definition encompasses all possible SGs within the quantum formalism. However, it is worth mentioning that with this embedding, the system does not exhibit quantum coherence since the density operator is always diagonal in the basis $\ket i \bra i$, as is readily seen by checking $\bra{i}\mathcal K_\symb  \rho\ket{j}\propto \delta_{ij}$ for any $\rho$. We now look at a class of SGs that always admit a coherent representation in terms of pure states and operations. These will provide the grounds for exploring dynamics not available in the classical formalism.\\*

\noindent \textbf{Definition} [\cite{wiesner_computation_2008}]: A \emph{deterministic} generator is one such that for each outcome $s$ and each internal state $j$, the state can only transit to another state $i$. This means that for each column $j$ in each matrix $T_\symb$ there is only one nonzero entry. Let $\Prob(\symb|j)$ be the only nonzero value in the $j$th column of $T_\symb$, and $I_j(\symb)$ be the row in which $\Prob(\symb|j)$ appears. The function $I$ will be useful in the following.

The idea behind deterministic generators is that given the initial state and the sequence of outcomes, the state of the system can always be tracked.\\*

\noindent \textbf{Definition:} A \emph{reversible} generator is a deterministic SG for which each $T_\symb$ has no more than one nonzero entry per row. Using the $I_j(\symb)$ notation, this reads
\begin{equation}
	\label{eq:reversible}
	I_j(\symb)=I_{j'}(\symb)\Longleftrightarrow j=j',
\end{equation}
or equivalently, $\delta_{I_j(\symb)\,I_{j'}(\symb)}=\delta_{jj'}$.
\\*

Reversible generators have the property that the state of the system can be tracked back if the final state and the sequence of outcomes are known.
\\*

\noindent \textbf{Theorem:} For reversible generators it is possible to define a \HQM ~in terms of pure operations.\\ 
\noindent \emph{Proof:} Let $\mathcal K_s$ be the quantum operation associated to symbol $s$, with operator sum representation 
\begin{equation}
	\mathcal K_s\rho=K_s\rho K_s^\dagger,\qquad K_\symb=\sum_{j}\sqrt{\Prob(\symb|j)}\ket{I_j(\symb)}\bra{j}.
\end{equation}
The unnormalized quantum state after outcome $\symb$ is
\begin{equation}
	K_\symb\ket{j}=\sqrt{\Prob(\symb|j)}\ket{I_j(\symb)},
\end{equation}
and outcome $s$ occurs with probability $\left|K_\symb\ket{j}\right|^2=\Prob(\symb|j)$. Finally, the trace-preserving condition is satisfied
\begin{equation}
	\sum_\symb  K^\dagger_\symb K_\symb=\sum_\symb \sum_{jj'}\delta_{jj'}\sqrt{\Prob(\symb|j)\Prob(\symb|j')}\ket{{j}}\bra{{j'}}
	=\sum_{j}\sum_\symb {\Prob(\symb|j)}\ket j\bra j=\openone,
\end{equation}
where the first equality follows by virtue of Eq.~(\ref{eq:reversible}).~$\blacksquare$

The set of SG that admit a \HQM ~representation in terms of pure states and operations is not restricted to reversible generators, but \emph{all} reversible generators admit such representation.

In analogy to Eq.~(\ref{eq:ssSG}), the stationary state $\rho^\stat$ is defined as
\begin{equation}
	\label{eq:ss}
	\rho^\stat=\sum_\symb \mathcal K_\symb \rho^\stat.
\end{equation}
and word probabilities are computed as
\begin{equation} \label{eq:wp}
	\Prob(\symb_1\ldots\symb_n)=\tr[\mathcal K_{\symb_n}\cdots \mathcal K_{\symb_1} \rho^\stat].
\end{equation}
Finally, it is worth noticing that defining the Hilbert-Schmidt product $(A|B)=\tr[A^\dagger B]$, $\Prob(\symb_1\ldots\symb_n)$ can be written as
\begin{equation} \label{eq:wpxxx}
	\Prob(\symb_1\ldots\symb_n)=(\openone|\mathcal K_{\symb_n}\cdots \mathcal K_{\symb_1} |\rho^\stat),
\end{equation}
in full analogy with Eq.~(\ref{eq:wpSG}).

\section{Non-adaptive read-out of many-body states} \label{sec:Cluster-state}

We now point out the relation between a Hidden Quantum Model and the
sequential readout of a many-body state. As an example we start with the
sequential readout of a 1D cluster state. As we shall see below, these
read-outs are examples of {\HQM}s which generate stochastic languages with
potentially long-range correlations. Although this particular example (1D
cluster states) does not provide universal resources for quantum computation,
the generalization to 2D cluster states, or, equivalently multiqubit {\HQM}s does, in principle, capture the power of quantum computation. 

\subsection{Preparation and non-adaptive read of 1D cluster states}

Let us start with a short review of 1D cluster states \cite{Briegel}. One way to generate a 1D cluster state of $N$ qubits is to first prepare $N$ qubits in the same state $|+ \rangle$,
\begin{eqnarray}
|+ \rangle \equiv \big( \ket0+\ket1 \big)/\sqrt{2} \, ,
\end{eqnarray}
which is an equal superposition of $|0 \rangle$ and $|1 \rangle$. Afterwards, $N-1$ controlled two-qubit phase gates $U_{n \, n-1}$, 
\begin{eqnarray} \nonumber
U_{n \, n-1} &\equiv & |00\rangle_{n \, n-1} \langle 00| +  |01\rangle_{n \, n-1} \langle 01|  +  |10\rangle_{n \, n-1} \langle 10|  -  |11 \rangle_{n \, n-1} \langle 11|\\
\label{Vij}
&=&\ket0_n\bra0\otimes\openone_{n-1}+\ket1_n\bra1\otimes Z_{n-1}\, , 
\end{eqnarray}
should be applied to all neighboring qubits. An $N$-qubit 1D cluster state can hence be written as
\begin{eqnarray} \label{clusterN}
\ket{{\rm cluster}_N} = U_{N\,N-1}\, \cdots \, U_{32} \, U_{21} \,
\ket{+}^{\otimes N} \, .
\end{eqnarray}

Once a large cluster state is prepared, the easiest way to generate a sequence of 0's and 1's is to perform successive single-qubit measurements. Suppose, a 0 is obtained, if the qubit is found in $\ket{\basis_0}$ and output 1 is obtained, if the qubit is found in $\ket{\basis_1}$ with
\begin{eqnarray} \label{phi0}
\ket{\basis_0} &\equiv & \cos \pangle \, \ket{0} + {\rm e}^{{\rm i} \xi} \, \sin \pangle \, \ket{1} \, , \nonumber \\
\ket{\basis_1} &\equiv & \sin \pangle \, \ket{0} - {\rm e}^{{\rm i} \xi} \, \cos \pangle \, \ket{1} 
\end{eqnarray}
with $\pangle$ and $\xi$ being free parameters. The corresponding projection
operators are $P_{s_i} = |\basis_{s_i} \rangle \langle \basis_{s_i}|$. For
example, $n$ measurements on a cluster state consisting of $N>n$ qubits
project them into the (unnormalised) state $P_{s_n} \cdots P_{s_1} \,
\ket{{\rm cluster}_N}$, where $s_i$ is the outcome of the $i$'th measurement,
and it is implicitly understood that $P_{s_i}$ acts on the $i$'th qubit. The
probability to obtain the word $\textbf{s}$ of length $n$ hence equals
\begin{eqnarray} \label{prob}
\Prob(\textbf{s}) &=& \| \, P_{s_n} \cdots P_{s_1} \, \ket{{\rm cluster}_N} \, \|^2 \, .
\end{eqnarray}
As we shall see below, the obtained bit sequences are completely random for certain parameters $\pangle$ and $\xi$ and contain long-range correlations for others.

\subsection{Non-adaptive cluster-state read-out in operator-sum representation} \label{sec.cluster-kraus}

\begin{figure}[t]
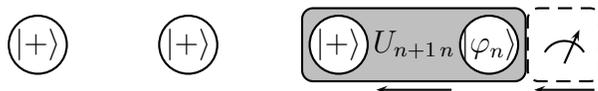

\newcommand{\atomplus}[1]{\pscircle[fillstyle=solid,fillcolor=white](#1,.5){.4} \rput(#1,.5){$\ket+$}}
\newcommand{\atom}[1]{\pscircle[fillstyle=solid,fillcolor=white](#1,.5){.4}}
\def\zigzag{\pscurve[linewidth=2pt](0,0)(0.1,0.1)(0.2,-0.1)(0.3,0)}
\begin{center}
	\pspicture(8,1)
	\psframe[framearc=.3,fillstyle=solid,fillcolor=lightgray](4,0)(7,1)
	\rput(5.5,.5){$U_{n+1 \, n}$}
	\psframe[framearc=.3,fillstyle=solid,fillcolor=white,linestyle=dashed](7,0)(8,1)
	\psarc(7.5,.25){.3}{30}{150}
	\psline{->}(7.5,.25)(7.7,.7)
	\psline{->}(6,-.1)(5,-.1)
	\psline{->}(7.9,-.1)(7.1,-.1)
	\atomplus{0.5}
	\atomplus{2.5}
	\atomplus{4.5}
	\atom{6.5}
	\rput(6.5,.5){$\ket{\varphi_n}$}
	\endpspicture
\end{center}
\caption{Sequential generation  and non-adaptive read-out of a one-dimensional cluster-state. Initially, all atoms are prepared in $|+ \rangle$. Afterwards, entangling phase gates $U_{n+1 \, n}$ are applied successively just before a measurement is performed on the state $|\varphi_n \rangle$ of qubit $n$.} \label{fig1}
\end{figure}

Before analysing the above process in more detail, let us show that the non-adaptive read-out of a cluster state is a concrete example of a \HQM ~with one qubit. The reason for this is that the entangling operations $U_{n \, n-1}$ for the preparation of the cluster state commute with the projections $P_{s_i}$ whenever $i$ is different from $n$ and $n-1$. It is therefore not necessary to complete the build up of a cluster state of size $N$ before measuring its qubits. Instead, one can constantly add new qubits to the chain, as illustrated in Fig.~\ref{fig1}, as long as $U_{n \, n-1}$ and $U_{n+1 \, n}$ are performed {\em before} the read-out of qubit $n$ takes place. 

To describe the cluster state read-out of qubit $n$, it is hence sufficient to
consider a chain of no more than $n$ qubits. After measuring the first $n-2$
qubits, the unnormalised  state of this chain equals 
\begin{eqnarray} \label{word1}
|+ \rangle \ket{\varphi(\textbf{s}^{n-2})}_{n-1} |\basis_{s_{n-2}}\rangle
\ldots |\basis_{s_1} \rangle = P_{s_{n-2}} U_{n-1 \, n-2} \, \cdots \, P_{s_1}
U_{21} \, \ket{+}^{\otimes N} 
\end{eqnarray}
Here $\ket{\varphi(\textbf{s}^{n-2})}_{n-1}$ denotes the state of qubit $n-1$
prior to its measurement and after measuring qubits $1$ to $n-2$ and obtaining
the word $\textbf{s}^{n-2}=(s_1,\ldots,s_{n-2})$. The state needs to be
normalised by  $\Prob({\bf s}^{n-1})$, obtained
from Eq.~\ref{prob}. Analogously, one can show that the (again unnormalised) state of the $n$ qubits after the detection of qubit $n-1$ is in this scenario given by 
\begin{eqnarray} \label{word}
\ket{\varphi(\textbf{s}^{n-1})}_n |\basis_{s_{n-1}}\rangle \ldots |\basis_{s_{1}} \rangle 
&=& P_{s_{n-1}} U_{n \, n-1} \, \cdots \, P_{s_1} U_{21} \, \ket{+}^{\otimes N} \, . 
\end{eqnarray}
with $\ket{\varphi(\textbf{s}^{n-1})}_n$ describing qubit $n$ prior to its
measurement. A comparison of these two equations then leads us to the relation
\begin{eqnarray} \label{word2}
\ket{\varphi(\textbf{s}^{n-1})}_n |\basis_{s_{n-1}} \rangle &= & P_{s_{n-1}}
U_{n \, n-1} \ket{+} \ket{\varphi(\textbf{s}^{n-2})}_{n-1} \, .
\end{eqnarray}
Using Eqs.~(\ref{Vij}) and applying $\openone\otimes\bra{\epsilon_{s_{n-1}}}$ from the left, this equation leads us to
\begin{eqnarray} \label{fire}
|\varphi(\textbf{s}^{n-1}) \rangle_n &= & K_{s_{n-1}} \, |\varphi(\textbf{s}^{n-2})\rangle_{n-1}
\end{eqnarray}
with  
\begin{eqnarray} \label{Kraus}
K_{s_i} =  \frac{1}{\sqrt{2}}\big( \, |0 \rangle \langle \basis_{s_i} | +  |1
\rangle \langle \basis_{s_i} | \, Z \, \big) 
~~ {\rm and} ~~ Z \equiv \ket0\bra0-\ket1\bra1 
\end{eqnarray}
which relates the state of qubit $n$ prior to its measurement to the state of qubit $n-1$ prior to its measurement. Analyzing the probabilities we see from Eq.~(\ref{word}) that $\Prob({\bf s}^{n-1})=\| \ket{\varphi(\textbf{s}^{n-1})}\|^2$, while Eq.~(\ref{fire}) can be written as
\begin{equation}
	\Prob({\bf s}_{n-1})=\left| K_{s_{n-1}} \, \frac{|\varphi(\textbf{s}^{n-2})\rangle}{\sqrt{\Prob(\textbf{s}^{n-2})}}\right|^2\Prob(\textbf{s}^{n-2}),
\end{equation}
showing that $\Prob(s_{n-1}|{\bf s}^{n-2})$ can be computed from the normalized state vector $|\varphi(\textbf{s}^{n-2})\rangle/{\sqrt{\Prob(\textbf{s}^{n-2})}}$.

From Eq.~(\ref{fire}) we see that the non-adaptive read-out of a 1D cluster state is equivalent to performing the same pure quantum operation over and over again on a single qubit. After $n-1$ measurements generating the word $\textbf{s}^{n-1}$ the state of this qubit equals (up to normalisation)
\begin{eqnarray}
\ket{\varphi(\textbf{s}^{n-1})} &=& K_{s_{n-1}} \cdots \, K_{s_1} \, |+ \rangle \, .
\end{eqnarray}
Here, we dropped the subscript $n$ in $\ket{\varphi(\textbf{s}^{n-1})}_n$
since we only describe a single qubit, but after $n-1$ time steps.
In other words, the non-adaptive read-out of a cluster state is a concrete
realisation of a \HQM~where the hidden quantum system consists of just one
qubit. The operators $K_0$ and $K_1$ indeed obey condition (\ref{eq.o-sum}) for
an operator-sum representation. The state $\ket+$ in Eq.~(\ref{word}) can be regarded as an ancilla which assists the detection of the qubit. After each measurement, the ancilla assumes the role of the qubit and the former qubit is replaced by a fresh ancilla prepared in $|+ \rangle$. This process is equivalent to {\em resetting} part of the system.  

Finally, let us comment on the information contained in the state $|\varphi(\textbf{s}^{n-1}) \rangle$ about previous measurement outcomes. Suppose $|\varphi(\textbf{s}^{n-2}) \rangle = \alpha \, |0 \rangle + \beta \, |1 \rangle$. Using Eqs.~(\ref{phi0}), (\ref{fire}), and (\ref{Kraus}) and calculate $ |\varphi(\textbf{s}^{n-1}) \rangle$ as a function of the complex coefficients $\alpha$ and $\beta$, we then find that
\begin{eqnarray} \label{crucial}
K_0 \, |\varphi(\textbf{s}^{n-2}) \rangle &= & \alpha \, \cos \pangle \, |+ \rangle + \beta \, {\rm e}^{-{\rm i} \xi} \sin \pangle \, |- \rangle \, , \nonumber \\
K_1 \, |\varphi(\textbf{s}^{n-2}) \rangle &= & \alpha \, \sin \pangle \, |+ \rangle - \beta \, {\rm e}^{-{\rm i} \xi} \cos \pangle \, |- \rangle 
\end{eqnarray}
with $|-\rangle \equiv \big( \ket0-\ket1 \big)/\sqrt{2}$. Both states depend on $\alpha$ and $\beta$, even after normalisation. This is different to von Neumann measurements which erase the coefficients $\alpha$ and $\beta$ when the respective projector is applied. In a \HQM, quantum information about the history of the system can in principle survive for many time steps. 

\subsection{Analysis of the language generated by non-adaptive cluster state read-out} \label{anna}

As suggested in Section \ref{properties}, we now calculate the stationary state $\rho^\stat$ of the non-adaptive cluster state read-out. Using Eq.~(\ref{eq:ss}) and the Kraus operators in Eq.~(\ref{Kraus}) we find that the stationary state of the system is a completely mixed state,
\begin{eqnarray}
	\rho^\stat={1 \over 2} \, \openone \, ,
\end{eqnarray}
independent of the parameters $\pangle$ and $\xi$ in Eq.~(\ref{phi0}). This state can now be used to predict word probabilities.  Using Eq.~(\ref{word}), we can easily show that the symbols 0 and 1 occur with the same frequency, i.e.~$\Prob(0) = \Prob (1) = {1 \over 2}$. Using Eqs.~(\ref{eq:wp}) and (\ref{eq:wpxxx}), we moreover find that all length-2 words occur equally often, \emph{i.e.} 
\begin{eqnarray}
	\Prob(00) = \Prob (01) = \Prob(10) = \Prob(11)={1\over 4}  \, .
\end{eqnarray}
This shows that the language created during the non-adaptive cluster state read out contains {\em no} length-2 correlations in the long run. However, looking at length-3 word probabilities, we find 
\begin{eqnarray}
	\Prob(000) &=& {1\over 8} +  {1 \over 32} \, \cos \xi \, \left( \sin 2 \pangle + \sin 6 \pangle \right) \, , \nonumber \\
	\Prob(111) &=& {1\over 8} -  {1 \over 32} \, \cos \xi \, \left( \sin 2 \pangle + \sin 6 \pangle \right) 
\end{eqnarray}
and
\begin{eqnarray}
	 \Prob(011) = \Prob(101) = \Prob(110) = \Prob(000) \, , \nonumber \\
	\Prob(001) = \Prob(010) = \Prob(100) = \Prob(111) \, . 
\end{eqnarray}
Depending on $\xi$ and $\pangle$, words with an even number of 1's can occur more often (or less often) than words with an odd number of 1's.

\begin{figure}
	\center\includegraphics{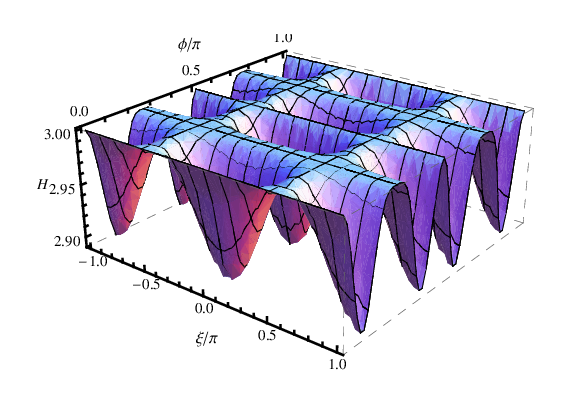}
	\caption{\label{fig:plot}Entropy of three-letter words ($H$ in bits) as a function of the measurement parameters $\xi$ and $\phi$. The entropy attains the maximum 3 for $\phi=k\, \pi/4$.}
\end{figure}

Looking at length-3 words we find that they provide an entropy 
\begin{eqnarray}
	\nonumber
	H_3&=&5-\frac{1}{2}\log\left(16-\cos^2\xi(\sin2\phi+\sin6\phi)^2\right)\\
	&&+\frac{1}{2}\cos\xi\cos^22\phi\sin2\phi\log\left(\frac{8}{4+\cos\xi(\sin2\phi+\sin6\phi)}-1\right)
\end{eqnarray}
which is smaller than 3 for values of $\xi\neq0$ and $\phi\neq k{\pi\over2}$, showing that there is length-3 correlation [see Fig.~\ref{fig:plot}].

\subsection{Sequential readout of matrix-product states}

\label{sec:MPS}

In this section we show that the statistics of sequential measurements on some
matrix-product-states (MPS) (see \cite{perez-garcia_matrix_2006} and
references therein) can be simulated by \HQM. More specifically, we concentrate on translationally-invariant \emph{open boundary condition} MPS with bond dimension $D$, of a many-body state $\ket\psi\in\mathcal H_{\mathcal B}^{\otimes n}$ of the form
\begin{equation}
	\label{eq:mps}
	\ket{\psi}=\sum_{i_1,\ldots,i_n}V_{[n]}^{i_n}\cdots V_{[1]}^{i_1}\ket{i_n,\ldots,i_1},
\end{equation}
where $i_k$ runs from 1 to $d=\dim \mathcal H_{\mathcal B}$ and $V^i_{[k]},~1<k<n$ are $D\times D$ matrices and the first and last are row and column vectors, respectively. In~\cite{schon_sequential_2005, perez-garcia_matrix_2006} it is shown that every MPS of the form of Eq.~{\ref{eq:mps}) can be generated sequentially by isometries $\mathcal H_{\mathcal A}\rightarrow \mathcal H_{\mathcal A}\otimes\mathcal H_{\mathcal B}$ of dimension $dD\times D$, with the constraint that the last isometry decouples the spin chain $\mathcal H_{\mathcal B}^{\otimes n}$ from the system $\mathcal H_{\mathcal A}$. The isometries are of the form 
\begin{equation}
	V=\sum_{i,\alpha,\beta} V^i_{\alpha,\beta}\ket{\alpha,i}\bra{\beta},
\end{equation}
and the assuming that $\ket\psi=\bra{\psi_F}\tilde V_{[n]}\cdots \tilde V_{[1]}\ket{\psi_I}$. Here greek indices refer to dimensions on system $\mathcal A$ and latin indices refer to system $\mathcal B$, and $\ket{\psi_I}$ and $\ket{\psi_F}$ are initial and final states of system $\mathcal A$, respectively. Since we are interested in generating stochastic processes with arbitrarily long sequences, we will only be concerned with the initial state. The final state will not enter our discussion. Moreover, translational invariance of the MPS representation means that all matrices $V^i_{[k]}$ are independent of the site index, $k$. We therefore we drop it from now on~\footnote{Notice that translational invariance of a quantum state does not necessarily mean that it has a representation with the same invariance.}. 

A more physically intuitive picture arises by considering the isometries as entangling unitary transformations $\mathcal V$ between system $\mathcal A$ and system  $\mathcal B$ in an initialized state $\ket0\in\mathcal H_{\mathcal B}$, such that when acting on an initialized ancilla yields
\begin{equation}
	\mathcal V\ket0=\sum_{i\alpha\beta}V^i_{\alpha,\beta}\ket{\alpha,i}\bra{\beta}.
\end{equation}

A note about semantics is in order. In the original paper~\cite{schon_sequential_2005} the term ancilla refers to system $\mathcal A$ whereas in our language we consider system $\mathcal B$ to be the ancilla. This is because in the literature, concern is focused in generating multipartite quantum states, and system $\mathcal A$ only assists in this process. Instead, in our approach, system $\mathcal B$ is considered only as an assistance for implementing generalized measurements on $\mathcal A$, hence we call $\mathcal B$ the \emph{ancilla}.

Now, after $k$th ancilla $\mathcal B$ has been involved in the corresponding isometry, it can be measured, just like in the example of Sec.~\ref{sec.cluster-kraus}. Let $\rho$ be the state of the system prior to step $k$. Attaching the initialized ancilla and performing the entangling operation yields $\mathcal V\,(\rho_{k-1}\otimes\ket0\bra0)\,\mathcal V^\dagger$. Now, performing a projective measurement on system $\mathcal B$ yields 
\begin{equation}
	\label{eq:MPSmeasurement}
	\rho_s=\tr_{\mathcal B}\big[(\openone\otimes P_s) \mathcal V \,(\rho\otimes\ket0\bra0)\,\mathcal V^\dagger (\openone\otimes P_s)\big]
\end{equation}
By defining the quantum operation $\mathcal K_s\cdot=\sum_i K_s^i\cdot K_s^i{}^\dagger$ with
\begin{equation}
	K_s^i=\bra{i}P_s \mathcal V\ket 0=\sum_{\alpha\beta j}\bra{i}P_s\ket{j}\,[V]^j_{\alpha\beta}\,\ket{\alpha}\bra{\beta},
\end{equation}
we can write
\begin{equation}
	\rho_s=\mathcal K_s \rho,
\end{equation}
while 
\begin{equation}
	\sum_{i,s}K_s^i{}^\dagger K_s^i=\openone,
\end{equation}
showing that $\{\mathcal K_s\}$ a well defined stochastic quantum operation.

This shows that, although arbitrary MPS states will require specific initial
and final isometries (initialization of system $\ket{\psi_I}$ and decoupling
in the last step, $\ket{\psi_F}$), we are only interested in long sequences of
outcomes for arbitrarily large MPS, and hence we can disregard the final
decoupling step. Therefore {\HQM}s can generate the same measurement statistics as sequential identical measurements on any many-body state that admits a translationally invariant MPS representation
\begin{equation}
	\ket\psi=\sum_{i_1\ldots i_n}\bra{\psi_F}V^{i_n}\cdots V^{i_1}\ket{\psi_I}\,\ket{i_n\ldots i_1} \, .
\end{equation}
The number of internal states of the required \HQM corresponds to the bond
dimension $D$ of such MPS representation. The study of random processes
generated by {\HQM}s should then be highly connected to the results on correlation lengths and other properties of certain many-body systems. To the best of our knowledge, this establishes a --yet unexplored-- two-way communication between the theory of many-body states and the theory of stochastic processes.

\section{An example where quantum outperforms classical} \label{variation}

After showing that {\HQM}s provide a convenient framework to model stochastic processes generated by a variety of quantum systems, it is reasonable to ask whether a more efficient description of a given stochastic process is allowed in the quantum formalism as compared to the classical case. A general answer to this question is beyond the scope of this paper. However, we will provide evidence that some stochastic processes admit a simpler representation within the quantum formalism. For this we present a 4-state classical model, and show that it cannot be reduced to fewer than three states. Then we provide a 2-state \HQM~that generates the same process.

Consider the 4-symbol stochastic process ($s\in\{0,1,2,3\}$) given by
(states labeled in the order $U, D, R, L$)
\begin{eqnarray}\label{eq:transfer_matrix}
T_0&=\left({\small
\begin{array}{cccc}
 1/2 & 0 & 1/4 & 1/4\\
 0   & 0 & 0 & 0\\
 0 & 0 & 0 & 0\\
 0 & 0 & 0 & 0
\end{array}}\right),
\quad
&T_1=\left({\small
\begin{array}{cccc}
 0 & 0 & 0 & 0 \\
 0 & 1/2 & 1/4 & 1/4\\
 0 & 0 & 0 & 0\\
 0 & 0 & 0 & 0
\end{array}}
\right) \, , \nonumber \\
T_2&=\left({\small
\begin{array}{cccc}
 0 & 0 & 0 & 0 \\
 0 & 0 & 0 & 0\\
 1/4 & 1/4 & 1/2 & 0\\
 0 & 0 & 0 & 0
\end{array}}
\right),
\quad
&T_3=\left({\small
\begin{array}{cccc}
 0 & 0 & 0 & 0 \\
 0 & 0 & 0 & 0\\
 0 & 0 & 0 & 0\\
 1/4 & 1/4 & 0 & 1/2
\end{array}}
\right).~~~~~~
\end{eqnarray}
and is depicted in Fig.~\ref{fig:diag}. 

\begin{figure}[t]
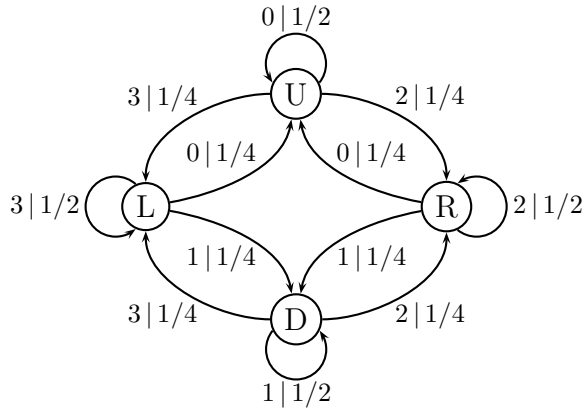

\begin{center}
	\pspicture(8,6)
	\psset{radius=.5,labelsep=0.0}
	%\psgrid
	\cnodeput(4,4.5){up}{U}
	\cnodeput(6,3){right}{R}
	\cnodeput(2,3){left}{L}
	\cnodeput(4,1.5){down}{D}

	\nccurve[angleA=0,angleB=90]{->}{up}{right}
	\naput{\footnotesize$2\,|\,1/4$}
	\nccurve[angleA=180,angleB=90]{->}{up}{left}
	\nbput{\footnotesize $3\,|\,1/4$}
	\nccircle[]{->}{up}{.4}
	\nbput{\footnotesize $0\,|\,1/2$}
	
	\nccurve[angleA=0,angleB=-90]{->}{down}{right}
	\nbput{\footnotesize$2\,|\,1/4$}
	\nccurve[angleA=180,angleB=-90]{->}{down}{left}
	\naput{\footnotesize $3\,|\,1/4$}
	\nccircle[angleA=180]{->}{down}{.4}
	\nbput{\footnotesize $1\,|\,1/2$}

	\nccurve[angleA=170,angleB=-80]{->}{right}{up}
	\nbput{\footnotesize$0\,|\,1/4$}
	\nccurve[angleA=190,angleB=80]{->}{right}{down}
	\naput{\footnotesize $1\,|\,1/4$}
	\nccircle[angleA=-90]{->}{right}{.4}
	\nbput{\footnotesize $\,2\,|\,1/2$}

	\nccurve[angleA=10,angleB=-100]{->}{left}{up}
	\naput{\footnotesize$0\,|\,1/4$}
	\nccurve[angleA=-10,angleB=100]{->}{left}{down}
	\nbput{\footnotesize $1\,|\,1/4$}
	\nccircle[angleA=90]{->}{left}{.4}
	\nbput{\footnotesize $3\,|\,1/2\,$}
	\endpspicture
\end{center}
\caption{Diagrammatic representation of the stochastic generator for the \HQM~defined by (\ref{eq:transfer_matrix}).}
\label{fig:diag}
\end{figure}

A simple method for obtaining a lower bound to the number of internal states necessary for representing a given HMM was provided in \cite{anderson_realization_1999}. This bound may turn out to be too loose, since positivity and other conditions are not enforced. However, it will suffice to make our point. In our argument we follow closely the approach in \cite{anderson_realization_1999}. Suppose the statistics of the model (\ref{eq:transfer_matrix}) can be obtained from a HMM with $n$ internal states. Then, consider the matrix
\begin{equation}\label{eq:hankel}
	H=\left(
		\begin{array}{lllll}
	 		\Prob(\emptyset) & \Prob(0) & \Prob(1) & \Prob(2) & \Prob(3) \\
	 		\Prob(0) & \Prob(00) & \Prob(10) & \Prob(20) & \Prob(30) \\
	 		\Prob(1) & \Prob(01) & \Prob(11) & \Prob(21) & \Prob(31) \\
	 		\Prob(2) & \Prob(02) & \Prob(12) & \Prob(22) & \Prob(32) \\
	 		\Prob(3) & \Prob(03) & \Prob(13) & \Prob(23) & \Prob(33)
		\end{array}
	\right).
\end{equation}
This is the upper-left corner of an infinite matrix known in the literature as the Hankel matrix. By convention it is assumed that $\Prob(\emptyset)=1$. Expanding the structure of $\Prob(s_1s_2)=\bra1 T_{s_2}T_{s_1}\ket\pi$ we can write
\begin{equation}\label{eq:hankelfactorized}
	H=\left(
		\begin{array}{l}
			\bra1T_\emptyset\\
			\bra1T_0\\
			\bra1T_1\\
			\bra1T_2\\
			\bra1T_3\\
		\end{array}
	\right)
	\left(
		\begin{array}{lllll}
			T_\emptyset\ket\pi&T_0\ket\pi&T_1\ket\pi&T_2\ket\pi&T_3\ket\pi
		\end{array}
	\right),
\end{equation}
where $T_\emptyset=\openone$. Notice that each block in this factorization is a vector of dimension $n$. Therefore the rank of both the left and the right matrices in Eq.~(\ref{eq:hankelfactorized}) are bounded by $n$. Since $H$ and all similar subblocks are factorized in the same manner, any sublock of the Hankel matrix, and in particular that shown in Eq.~(\ref{eq:hankel}), cannot have rank higher than $n$. Assuming that $n$ could be reduced to 2, we are thus led to the conclusion that $\rank(H)\leq2$. It is now straightforward to explicitly compute $\rank(H)$. For this, use Eq.~(\ref{eq:transfer_matrix}) together with $\sum_s T_s\ket\pi=\ket\pi$, with $
\ket\pi=\frac{1}{4}\ket1$. Then
\begin{equation}
	\rank(H)=\rank\left(
		\begin{array}{lllll}
			1 & \frac{1}{4} & \frac{1}{4} & \frac{1}{4} & \frac{1}{4} \\
			\frac{1}{4} & \frac{1}{8} & 0 & \frac{1}{16} & \frac{1}{16} \\
			\frac{1}{4} & 0 & \frac{1}{8} & \frac{1}{16} & \frac{1}{16} \\
			\frac{1}{4} & \frac{1}{16} & \frac{1}{16} & \frac{1}{8} & 0 \\
			\frac{1}{4} & \frac{1}{16} & \frac{1}{16} & 0 & \frac{1}{8}
		\end{array}
	\right)=3 \, .
\end{equation}
This shows that the model described in Eq.~(\ref{eq:transfer_matrix}) cannot
be implemented by a two-state HMM. It is, however, generated by a \HQM described by ($\mathcal K_y\cdot=K_y\cdot K_y^\dagger$):
	\begin{eqnarray} \label{eqs:hqm}
		K_{0}&=&\frac{1}{\sqrt2}\ket{\uparrow}\bra{\uparrow}, \qquad K_2=\frac{1}			{\sqrt2}\ket+\bra+ \, , \nonumber \\
		K_{1}&=&\frac{1}{\sqrt2}\ket{\downarrow}\bra{\downarrow}, \qquad K_3=				\frac{1}{\sqrt2}\ket-\bra- \, .
	\end{eqnarray}
Indeed, the HMM in Eq.~(\ref{eq:transfer_matrix}) can easily be derived from
the \HQM~in Eq.~(\ref{eqs:hqm}). 

\section{Discussion}\label{sec:conclusions}

We have introduced Hidden Quantum Models, inspired by the success of Hidden
Markov Models in modelling a wide variety of relevant stochastic processes,
and by the natural analogy that exists between the latter and the modern
formalism of Quantum Mechanics. We expect {\HQM}s to provide a wide framework
which allows to systematically study quantum processes in which the access to
the quantum system is limited, and measurements are constrained to some
constant and sequential leaking of classical information. We have considered
the least restrictive approach and considered most general CP maps which
describe not only the outcome statistics of the measurements but also the
effect on the quantum system. These models are a generalization to previously
considered classical automata, such as stochastic generators, Mealy automata
and hidden Markov models. At the same time, our models generalize the class of
quantum generators considered previously in the literature. The advantages of
considering {\HQM}s are many-fold. On the one hand, it is sufficiently general to naturally encompass all Hidden Markov Models, a highly desirable feature for any quantum model attempting to outperform their classical counterpart. On the other hand, they provide the most compressed description of a quantum process by only accounting for the dynamics of the relevant part of the system, \emph{i.e.}, that which contains the quantum information latent in the system.

We foresee a number of questions related to {\HQM}s with practical and
theoretical relevance. We have shown that {\HQM}s can provide a reduction of the
necessary number of internal states necessary for generating some stochastic
processes, as compared with classical generators. It is then natural to ask to
what extent does this reduction take place, \emph{i.e.}, whether a significant
reduction of the resources can take place, or classical generators only
require a polynomial overhead on the resources required by {\HQM}s. On a related
ground, {\HQM}s are fit to describe processes which are simple enough to be
considered repetitive, but have an unavoidable quantum essence. It is
interesting to ask whether the dimensionality of the quantum system induces
some limitations on the generated stochastic processes. This would certainly
have implications such as fundamental dimensionality tests for quantum systems
under limited accessing freedom. Moreover, in sight of the relation between
{\HQM}s and MPS, unraveling the capabilities and limitations of {\HQM}s may yield insight on the role of adaptivity in measurement-based quantum computation.

On a more applied level, it is worth noticing that a well established theory
of inference exists for HMMs~\cite{hastie_elements_2009, capp_inference_2007}.
This can be roughly divided in two classes, namely, \emph{state inference} and
\emph{parameter inference}. Both of these have close cousins in the field of
quantum metrology~\cite{helstrom_quantum_1976, holevo_statistical_2001}. It is
natural to ask to what extent can these tools be adapted for addressing
questions regarding {\HQM}s, such as those collected in~\cite{capp_inference_2007}. It is worth mentioning that some of these questions have already triggered interest within the quantum information community~\cite{juba_learning_2009}.\\*

\noindent{\em Acknowledgment.}\\*

The authors wish to thank useful discussions with E.~Bagan, J.~Calsamiglia, R.~Mu\~{n}oz-Tapia and M.~Plenio. A.M. acknowledges finnancial support from the European Commission of the European Union under the FP7 STREP Project HIP (Hybrid Information Processing), Grant Agreement n. 221889, from MIUR under the FARB fund, from INFN under Iniziativa Specifica PG 62, from CNR-INFM Center Coherentia and from the UK Engineering and Physical Sciences Research Council through the QIP IRC. A.B. thanks the Royal Society and the GCHQ for their support through a James Ellis University Research Fellowship. KW acknowledges funding through EPSRC grant EP/E501214/1. \\*
\vspace{.5cm}

\noindent{\em References}\\*

\end{document}